\documentclass[10pt, final, technote]{IEEEtran}
\usepackage{cite}

\ifCLASSINFOpdf
  \usepackage[pdftex]{graphicx}
\else
\fi
\usepackage[cmex10]{amsmath}
\usepackage{array}

\usepackage[tight,footnotesize]{subfigure}

\usepackage{fixltx2e}
\usepackage{url}

\usepackage[tight,ugly]{units}
\usepackage{mathtools}
\usepackage{algorithm}
\usepackage{algpseudocode}
\usepackage{color}
\usepackage{colortbl}
\usepackage{dsfont}
\usepackage{dblfloatfix}
\usepackage{flushend}

\newcommand{\R}[0]{\mathds{R}} 
\newcommand{\Z}[0]{\mathds{Z}} 
\newcommand{\gauss}[2]{\mathcal{N}(#1,#2)}

\newcommand{\gaussx}[3]{\mathcal{N}(#1\,|\,#2,#3)}
\newcommand{\gaussxBig}[3]{\mathcal{N}\left(#1\left|#2,#3\right.\right)}
\renewcommand{\d}{\operatorname{d}\!}
\newcommand{\E}{\mathds{E}} 
\newcommand{\var}{\mathrm{var}} 
\newcommand{\cov}[0]{\mathrm{cov}} 
\renewcommand{\vec}[1]{{\boldsymbol{\mathbf{#1}}}} 
\newcommand{\mat}[1]{{\ensuremath{\mathbf{#1}}}} 
\newcommand{\inv}[0]{^{-1}} 
\newcommand{\GP}[0]{\mathcal{GP}} 
\newcommand{\data}[0]{\mathcal{D}} 
\newcommand{\T}[0]{^\top} 
\newcommand{\prob}{{p}} 
\newcommand{\tr}{\mathrm{tr}} 
\newcommand{\diag}{\mathrm{diag}} 

\newcommand{\twofig}{0.403\hsize}
 
\newcommand{\idx}[1]{{(#1)}}

\newcommand{\mL}[0]{\mat Q} 
\newcommand{\obs}[0]{z} 
\newcommand{\nll}[0]{\mathrm{NLL}} 
\newcommand{\rmse}[0]{\mathrm{RMSE}} 
\newcommand{\mae}[0]{\mathrm{MAE}} 

\definecolor{green}{rgb}{0,0.6,0}
\newcommand{\green}[1]{\color{green}#1}
\definecolor{red}{rgb}{0.6,0,0}
\newcommand{\red}[1]{\color{red}#1}
\newcommand{\gk}[0]{\mathcal{K}}
\newcommand{\argmax}{\operatornamewithlimits{argmax}}

\begin{document}
%
\title{Robust Filtering and Smoothing with Gaussian Processes}
%
%
%

\author{Marc Peter Deisenroth$^{a,b,\ast}$, Ryan Turner$^b$, Marco
  F.~Huber$^{c}$, \\Uwe D.~Hanebeck$^d$, Carl Edward
  Rasmussen$^{b,e}$
  \thanks{$^\ast$ Corresponding author} \thanks{$^a$ University of
    Washington, Seattle, WA, USA}
  \thanks{$^b$ University of Cambridge, UK}
  \thanks{$^c$ AGT Group (R\&D) GmbH, Darmstadt,
    Germany}
  \thanks{$^d$ Karlsruhe Institute of Technology,
    Germany}
  \thanks{$^e$ Max Planck Institute for Biological Cybernetics,
    T\"ubingen, Germany}

}

\maketitle

\begin{abstract}
  We propose a principled algorithm for robust Bayesian filtering and
  smoothing in nonlinear stochastic dynamic systems when both the
  transition function and the measurement function are described by
  non-parametric Gaussian process (GP) models. GPs are gaining
  increasing importance in signal processing, machine learning,
  robotics, and control for representing unknown system
  functions by posterior probability distributions. This modern way of
  ``system identification'' is more robust than finding point
  estimates of a parametric function representation. 
  In this article, we present a principled algorithm for robust
  analytic smoothing in GP dynamic systems, which are increasingly
  used in robotics and control.  Our numerical evaluations demonstrate
  the robustness of the proposed approach in situations where other
  state-of-the-art Gaussian filters and smoothers can fail.
\end{abstract}


%
\IEEEpeerreviewmaketitle


%
%
%
%

\section{Introduction}

Filtering and smoothing in context of dynamic systems refers to a
Bayesian methodology for computing posterior distributions of the
latent state based on a history of noisy measurements. This kind of
methodology can be found, e.g., in navigation, control engineering,
robotics, and machine
learning~\cite{Anderson2005,Astrom2006,Thrun2005,
  Roweis2001}. Solutions to filtering~\cite{Kalman1960, Anderson2005,
  Astrom2006,Thrun2005,Roweis2001} and
smoothing~\cite{Rauch1965, Roweis1999, Kschischang2001, Pearl1988} in
linear dynamic systems are well known, and numerous approximations for
nonlinear systems have been proposed, for both
filtering~\cite{Maybeck1979, Julier2004, Ko2009, Deisenroth2009a,
  Hanebeck2003b, Arasaratnam2009} and smoothing~\cite{Sarkka2008,
  Godsill2004, Kitagawa1996, Deisenroth2011}.

In this article, we focus on Gaussian filtering and smoothing in
Gaussian process (GP) dynamic systems. GPs are a robust non-parametric
method for approximating unknown functions by a posterior distribution
over them~\cite{MacKay1998, Rasmussen2006}. Although GPs have been
around for decades, they only recently became computationally
interesting for applications in robotics, control, and machine
learning~\cite{Nguyen-Tuong2009, Murray-Smith2003, Kocijan2003,
  Deisenroth2011b, Deisenroth2011c}.

The contribution of this article is the derivation of a novel,
principled, and robust Rauch-Tung-Striebel (RTS) smoother for GP
dynamic systems, which we call the \emph{GP-RTSS}\@. The GP-RTSS
computes a Gaussian approximation to the smoothing distribution in
closed form. The posterior filtering and smoothing distributions can
be computed \emph{without} linearization~\cite{Maybeck1979} or
(small) sampling approximations of densities~\cite{Julier2004}.

We provide numerical evidence that the GP-RTSS is more robust than
state-of-the-art nonlinear Gaussian filtering and smoothing algorithms
including the extended Kalman filter (EKF)~\cite{Maybeck1979}, the
unscented Kalman filter (UKF)~\cite{Julier2004}, the cubature Kalman
filter (CKF)~\cite{Arasaratnam2009}, the GP-UKF~\cite{Ko2009}, and
their corresponding RTS smoothers. 
\emph{Robustness} refers to the ability of an inferred distribution to
explain the ``true'' state\slash measurement.

The paper is structured as follows: In
Secs.~\ref{sec:notation}--\ref{sec:smoothing}, we introduce the
problem setup and necessary background on Gaussian smoothing and GP
dynamic systems. In Sec.~\ref{sec:gps}, we briefly introduce Gaussian
process regression, discuss the expressiveness of a GP, and explain
how to train GPs. Sec.~\ref{sec:gp smoothing} details our proposed
method (GP-RTSS) for smoothing in GP dynamic systems. In
Sec.~\ref{sec:results}, we provide experimental evidence of the
robustness of the GP-RTSS\@. Sec.~\ref{sec:discussion} concludes the
paper with a discussion.

\subsection{Problem Formulation and Notation}
\label{sec:notation}
In this article, we consider discrete-time stochastic systems
\begin{align}
\vec x_t &= f(\vec x_{t-1}) + \vec w_{t}\,,\label{eq:system equation}\\
\vec\obs_t& = g(\vec x_t) + \vec v_t\,,\label{eq:measurement}
\end{align}
where $\vec x_t\in\R^D$ is the state, $\vec\obs_t\in\R^E$ is the
measurement at time step $t$, $\vec w_t\sim\gauss{\vec
  0}{\mat\Sigma_w}$ is Gaussian system noise, $\vec v_t\sim\gauss{\vec
  0}{\mat\Sigma_v}$ is Gaussian measurement noise, $f$ is the
transition function (or system function) and $g$ is the measurement
function. The discrete time steps $t$ run from 0 to $T$. The initial
state $\vec x_0$ of the time series is distributed according to a
Gaussian prior distribution $\prob(\vec x_0) =
\gauss{\vec\mu_0^x}{\mat\Sigma_0^x}$. The purpose of filtering and
smoothing is to find approximations to the posterior distributions
$\prob(\vec x_t|\vec\obs_{1:\tau})$, where $1\!\!:\!\!\tau$ in a subindex
abbreviates $1,\dotsc,\tau$ with $\tau=t$ during filtering and
$\tau=T$ during smoothing. In this article, we consider Gaussian
approximations $\prob(\vec x_t|\vec\obs_{1:\tau})\approx \gaussx{\vec
  x_t}{\vec\mu_{t|\tau}^x}{\mat\Sigma_{t|\tau}^x}$ of the latent state
posterior distributions $\prob(\vec x_t|\vec\obs_{1:\tau})$. We use
the short-hand notation $\vec a_{b|c}^d$ where $\vec a=\vec\mu$
denotes the mean $\vec\mu$ and $ \vec a = \mat\Sigma$ denotes the
covariance, $b$ denotes the time step under consideration, $c$ denotes
the time step up to which we consider measurements, and
$d\in\{x,\obs\}$ denotes either the latent space ($x$) or the observed
space ($\obs$)\@.

\subsection{Gaussian RTS Smoothing}
\label{sec:smoothing}
Given the filtering distributions $\prob(\vec
x_t|\vec\obs_{1:t})=\gaussx{\vec
  x_t}{\vec\mu_{t|t}^x}{\mat\Sigma_{t|t}^x}$, $t = 1,\dotsc, T$, a
sufficient condition for Gaussian smoothing is the computation of
Gaussian approximations of the joint distributions $\prob(\vec
x_{t-1},\vec x_t|\vec\obs_{1:t-1})$, $t =
1,\dotsc,T$~\cite{Deisenroth2011}.

In Gaussian  smoothers, the standard smoothing distribution for the
dynamic system in Eqs.~(\ref{eq:system
  equation})--(\ref{eq:measurement}) is always
\begin{align}
  \prob(\vec x_{t-1}|\vec\obs_{1:T}) &= \gaussx{\vec
    x_{t-1}}{\vec\mu_{t-1|T}^x}{\mat\Sigma_{t-1|T}^x}\,, \quad\text{where}\\
  \vec\mu_{t-1|T}^x &= \vec\mu_{t-1|t-1}^x + \mat
  J_{t-1}(\vec\mu_{t|T} -
  \vec\mu_{t|t})\label{eq:smoother mean}\,,\\
  \mat\Sigma_{t-1|T}^x &= \mat\Sigma_{t-1|t-1} + \mat
  J_{t-1}(\mat\Sigma_{t|T} -
  \mat\Sigma_{t|t})\mat J_{t-1}\T\label{smoother cov}\,,\\
  \mat
  J_{t-1}&\coloneqq\mat\Sigma_{t-1,t|t-1}^x(\mat\Sigma_{t|t-1}^x)\inv
  \,,\quad t = T,\dotsc,1\,.\label{eq:J-matrix}
\end{align}
%
Depending on the methodology of computing this joint distribution, we
can directly derive arbitrary RTS smoothing algorithms, including the
URTSS~\cite{Sarkka2008}, the EKS~\cite{Maybeck1979, Anderson2005}, the
CKS~\cite{Deisenroth2011}, a smoothing extension to the
CKF~\cite{Arasaratnam2009}, or the GP-URTSS, a smoothing extension to
the GP-UKF~\cite{Ko2009}. The individual smoothers (URTSS, EKS, CKS,
GP-based smoothers etc.) simply differ in the way of computing\slash
estimating the means and covariances required in
Eqs.~(\ref{eq:smoother
  mean})--(\ref{eq:J-matrix})~\cite{Deisenroth2011}\@.

To derive the GP-URTSS, we closely follow the derivation of the
URTSS~\cite{Sarkka2008}. The GP-URTSS is a novel smoother, but its
derivation is relatively straightforward and therefore not detailed
in this article. 
Instead, we detail the derivation of the GP-RTSS, a robust
Rauch-Tung-Striebel smoother for GP dynamic systems, which is based on
analytic computation of the means and (cross-)covariances in
Eqs.~(\ref{eq:smoother mean})--(\ref{eq:J-matrix})\@.

In \emph{GP dynamics systems}, the transition function $f$ and the
measurement function $g$ in Eqs.~(\ref{eq:system
  equation})--\eqref{eq:measurement} are modeled by Gaussian
processes. This setup is getting more relevant in practical
applications such as robotics and control, where it can be difficult
to find an accurate parametric form of $f$ and $g$,
respectively~\cite{Atkeson1997a,Deisenroth2011b}. Given the
increasing use of GP models in robotics and control, the robustness of
Bayesian state estimation is important.

\section{Gaussian Processes}
\label{sec:gps}
In the standard GP regression model, we assume that the data
$\data\coloneqq\{\mat X\coloneqq[\vec x_1, \dotsc, \vec x_n]\T,\,\vec
y\coloneqq[y_1, \dots , y_n]\T\}$ have been generated according to
$y_i=h(\vec x_i) + \varepsilon_i$, where $h:\R^D\to\R$ and
$\varepsilon_i\sim\mathcal N(0,\sigma_\varepsilon^2)$ is independent
(measurement) noise. GPs consider $h$ a random function and infer a
posterior distribution over $h$ from data. The posterior is used to
make predictions about function values $h(\vec x_*)$ for arbitrary
inputs $\vec x_*\in\R^D$.

Similar to a Gaussian distribution, which is fully specified by a mean
vector and a covariance matrix, a GP is fully specified by a mean
\emph{function} $m_h(\,\cdot\,)$ and a covariance \emph{function}
\begin{align}
  k_h(\vec x,\vec x^\prime)&\coloneqq\E_h[(h(\vec x)-m_h(\vec x))(h(\vec
  x^\prime)-m_h(\vec x^\prime))]\\
  &=\cov_h[h(\vec x),h(\vec
  x^\prime)]\in\R\,,\quad\vec
  x,~\vec x^\prime\in\R^D\,,
\end{align}
which specifies the covariance between any two function values. Here,
$\E_h$ denotes the expectation with respect to the function $h$. The
covariance function $k_h(\,\cdot\,,\,\cdot\,)$ is also called a
\emph{kernel}.

Unless stated otherwise, we consider a prior mean function $m_h
\equiv 0$ and use the squared exponential (SE) covariance function
with automatic relevance determination
\begin{equation}\label{eq:SE kernel}
  k_{\text{SE}}(\vec x_p,\vec x_q) \coloneqq
  \alpha^2\exp\big(-\tfrac{1}{2}(\vec x_p-\vec
  x_q)\T\mat\Lambda\inv(\vec x_p-\vec
  x_q)\big)
\end{equation}
for $\vec x_p,\,\vec x_q\in\R^D$, plus a noise covariance
function $k_{\text{noise}}\coloneqq\delta_{pq}\sigma_\varepsilon^2$,
such that $k_h=k_{\text{SE}} + k_{\text{noise}}$. The $\delta$ denotes
the Kronecker symbol that is unity when $p=q$ and zero otherwise,
resulting in i.i.d.\ measurement noise. In Eq.~\eqref{eq:SE kernel},
$\mat\Lambda=\diag([\ell_1^2,\dotsc,\ell_D^2])$ is a diagonal matrix
of squared characteristic length-scales $\ell_i$, $i=1,\dotsc,D,$ and
$\alpha^2$ is the signal variance of the latent function $h$. By
using the SE covariance function from Eq.~(\ref{eq:SE kernel}) we
assume that the latent function $h$ is smooth and
stationary. Smoothness and stationarity is easier to justify than
fixed parametric form of the underlying function.


\subsection{Expressiveness of the Model}
Although the SE covariance function and the zero-prior mean function
are common defaults, they retain a great deal of expressiveness.
Inspired by~\cite{MacKay1998, Kern2000}, we demonstrate this
expressiveness and show the correspondence of our GP model to a
universal function approximator: Consider a function
\begin{align}
  \hspace{-2.3mm} h(x) =\sum\nolimits_{i\in\Z}\lim_{N\to\infty}
  \frac{1}{N}\sum\nolimits_{n = 1}^N
  \gamma_n\exp\left(-\frac{(x-(i+\tfrac{n}{N}))^2}{\lambda^2}\right)
  \label{eq:universal fct approx sum}
\end{align}
where $\gamma_n\sim\gauss{0}{1}\,, n = 1,\dotsc,N$. Note that in the
limit $h(x)$ is represented by infinitely many Gaussian-shaped basis
functions along the real axis with variance $\lambda^2$ and prior
(Gaussian) random weights $\gamma_n$, for $x\in\R$, and for all
$i\in\Z$. The model in Eq.~\eqref{eq:universal fct approx sum} is
considered a universal function approximator. Writing the sums in
Eq.~\eqref{eq:universal fct approx sum} as an integral over the real
axis $\R$, we obtain
\begin{align}
  h(x) &= \sum\nolimits_{i\in\Z} \int_i^{i+1}\gamma(s)
  \exp\left(-\frac{(x-s)^2}{\lambda^2}\right)\d s\nonumber\\
  &= \int_{-\infty}^\infty \gamma(s)
  \exp\left(-\frac{(x-s)^2}{\lambda^2}\right)\d s = (\gamma*\gk)(x)\,,
\label{eq:universal fct approx integral}
\end{align}
where $\gamma(s)\sim\gauss{0}{1}$ is a white-noise process and $\gk$
is a Gaussian convolution kernel.  The function values of $h$ are
jointly normal, which follows from the convolution $\gamma*\gk$. We
now analyze the mean function and the covariance function of $h$,
which fully specify the distribution of $h$. The only random variables
are the weights $\gamma(s)$.  Computing the expected function of this
model (prior mean function) requires averaging over $\gamma(s)$ and
yields
\begin{align}
  \E_\gamma[h(x)] &= \int h(x)\prob(\gamma(s))\d
  \gamma(s)\\
  &\hspace{-10mm}\stackrel{\eqref{eq:universal fct approx integral}}{=} \int
  \exp\left(-\frac{(x-s)^2}{\lambda^2}\right)\int\gamma(s)\prob(\gamma(s))\d
  \gamma(s)\d s = 0
\label{eq:universal fct approx meanFct}
\end{align}
since $\E_\gamma[\gamma(s)]=0$. Hence, the mean function of $h$ equals
zero everywhere. Let us now find the covariance function. Since the
mean function equals zero, for any $x,x^\prime\in\R$ we obtain
\begin{align}
  \cov_\gamma&[h(x), h(x^\prime)] = \int
  h(x)h(x^\prime)\prob(\gamma(s)) \d\gamma(s)\nonumber \\
  &= \int \exp\left(-\frac{(x-s)^2}{\lambda^2}\right)
  \exp\left(-\frac{(x^\prime-s)^2}{\lambda^2}\right)\nonumber\\
  &\quad\times \int\gamma(s)^2 \prob(\gamma(s))\d\gamma(s)\d s\,,
\end{align}
where we used the definition of $h$ in Eq.~\eqref{eq:universal fct
  approx integral}. Using $\var_\gamma[\gamma(s)] = 1$ and completing
the squares yields
\begin{align}
  \cov_\gamma[h(x), h(x^\prime)]
  &=\int\exp\left(-\frac{2\big(s-\tfrac{x+x^\prime}{2}\big)^2 +
      \tfrac{(x-x^\prime)^2}{2}}{\lambda^2} \right)\d s\nonumber\\
&= \alpha^2
  \exp\left(-\frac{(x-x^\prime)^2}{2\lambda^2}\right)
  \label{eq:universal fct approx covFct}
\end{align}
for suitable $\alpha^2$.

From Eqs.~\eqref{eq:universal fct approx meanFct}
and~\eqref{eq:universal fct approx covFct}, we see that the mean
function and the covariance function of the universal function
approximator in Eq.~\eqref{eq:universal fct approx sum} correspond to
the GP model assumptions we made earlier: a prior mean function
$m_h\equiv 0$ and the SE covariance function in Eq.~\eqref{eq:SE
  kernel} for a one-dimensional input space.
Hence, the considered GP prior implicitly assumes latent functions $h$
that can be described by the universal function approximator in
Eq.~\eqref{eq:universal fct approx integral}.
Examples of covariance functions that encode different model
assumptions are given in~\cite{Rasmussen2006}.

\subsection{Training via Evidence Maximization}
For $E$ target dimensions, we train $E$ GPs assuming that the target
dimensions are independent at a deterministically given test input (if
the test input is uncertain, the target dimensions covary): After
observing a data set $\data$, for each (training) target dimension, we
learn the $D+1$ hyper-parameters of the covariance function and the
noise variance of the data using \emph{evidence
  maximization}~\cite{MacKay1998,Rasmussen2006}: Collecting all
$(D+2)E$ hyper-parameters in the vector $\vec\theta$, evidence
maximization yields a point estimate
$\hat{\vec\theta}\in\argmax_{\vec\theta}\log\prob(\vec y|\mat
X,\vec\theta)$.  Evidence maximization automatically trades off data
fit with function complexity and avoids
overfitting~\cite{Rasmussen2006}.

From here onward, we consider the GP dynamics system setup, where two
GP models have been trained using evidence maximization: $\GP_f$,
which models the mapping $\vec x_{t-1}\mapsto\vec x_t,~\R^D\to\R^D$,
see Eq.~(\ref{eq:system equation}), and $\GP_g$, which models the
mapping $\vec x_t\mapsto\vec\obs_t,~\R^D\to\R^E$, see
Eq.~(\ref{eq:measurement})\@. To keep the notation uncluttered, we do
not explicitly condition on the hyper-parameters $\hat{\vec\theta}$
and the training data $\data$ in the following.

\section{Robust Smoothing in Gaussian Process Dynamic Systems}
\label{sec:gp smoothing}
Analytic moment-based filtering in GP dynamic systems has been
proposed in~\cite{Deisenroth2009a}, where the filter distribution is
given by
\begin{align}
\prob(\vec x_t|\vec\obs_{1:t}) &= \gaussx{\vec
  x_t}{\vec\mu_{t|t}^x}{\mat\Sigma_{t|t}^x}\,, \\
\vec\mu_{t|t}^x& = \vec\mu_{t|t-1}^x +
  \mat\Sigma_{t|t-1}^{x\obs}\big(\mat\Sigma_{t|t-1}^\obs\big)\inv
  (\vec\obs_t-\vec\mu_{t|t-1}^\obs)\,,\label{eq:generic filter mean}\\
  \mat\Sigma_{t|t}^x &= \mat\Sigma_{t|t-1}^x -
  \mat\Sigma_{t|t-1}^{x\obs}\big(\mat\Sigma_{t|t-1}^\obs\big)\inv
  \mat\Sigma_{t|t-1}^{\obs x}\label{eq:generic filter covariance}\,,
\end{align}
for $t = 1,\dotsc,T$.
Here, we extend these filtering results to analytic moment-based
smoothing, where we explicitly take nonlinearities into account (no
linearization required) while propagating full Gaussian densities (no
sigma/cubature-point representation required) through nonlinear GP
models.

In the following, we detail our novel RTS smoothing approach for GP
dynamic systems. We fit our smoother in the standard frame of
Eqs.~(\ref{eq:smoother mean})--(\ref{eq:J-matrix})\@. For this, we
compute the means and covariances of the Gaussian approximation
\begin{align}
\hspace{-1mm}
\gaussxBig{
\begin{bmatrix}
\vec x_{t-1}\\
\vec x_t
\end{bmatrix}
}{
\begin{bmatrix}
\vec\mu_{t-1|t-1}^x\\
\vec\mu_{t|t-1}^x
\end{bmatrix}
}{
\begin{bmatrix}
\mat\Sigma_{t-1|t-1}^x & \mat\Sigma_{t-1,t|t-1}^x\\
\mat\Sigma_{t,t-1|t-1}^x & \mat\Sigma_{t|t-1}^x
\end{bmatrix}
}
\label{eq:gp-ads desired joint}
\end{align}
to the joint $\prob(\vec x_{t-1},\vec x_t|\vec\obs_{1:t-1}) $, after
which the smoother is fully determined~\cite{Deisenroth2011}.  Our
approximation does not involve sampling, linearization, or numerical
integration. Instead, we present closed-form expressions of a
deterministic Gaussian approximation of the joint distribution in
Eq.~(\ref{eq:gp-ads desired joint})\@.

In our case, the mapping $\vec x_{t-1} \mapsto \vec x_t$ is not known, but
instead it is distributed according to $\GP_f$, a distribution over
system functions. For robust filtering and smoothing, we therefore
need to take the GP (model) uncertainty into account by Bayesian
averaging according to the GP
distribution~\cite{Quinonero-Candela2003a, Deisenroth2009a}. The
marginal $\prob(\vec
x_{t-1}|\vec\obs_{1:t-1})=\gauss{\vec\mu_{t-1|t-1}^x}{\mat\Sigma_{t-1|t-1}^x}$
is known from filtering~\cite{Deisenroth2009a}. In
Sec.~\ref{sec:marginal}, we compute the mean and covariance of second
marginal $\prob(\vec x_t|\vec\obs_{1:t-1})$ and then in
Sec.~\ref{sec:cross-covariance} the cross-covariance terms
$\mat\Sigma_{t-1,t|t-1}^x = \cov[\vec x_{t-1},\vec
x_t|\vec\obs_{1:t-1}]$.

\subsection{Marginal Distribution}
\label{sec:marginal}
\subsubsection{Marginal Mean}
Using the system Eq.~(\ref{eq:system equation}) and integrating
over all three sources of uncertainties (the system noise, the state $\vec
x_{t-1}$, and the system function itself), we apply the law of total
expectation and obtain the marginal mean
\begin{align}
  \vec\mu_{t|t-1}^x & =\E_{\vec x_{t-1}}\big[\E_f[f(\vec x_{t-1})|\vec
  x_{t-1}]|\vec\obs_{1:t-1}\big]\,.
\label{eq:gp-adf marginal mean 1st attempt}
\end{align}
The expectations in Eq.~(\ref{eq:gp-adf marginal mean 1st attempt})
are taken with respect to the posterior GP distribution $\prob(f)$ and
the filter distribution $\prob(\vec x_{t-1}|\vec\obs_{1:t-1})=
\gauss{\vec\mu_{t-1|t-1}^x}{\mat\Sigma_{t-1|t-1}^x}$ at time step
$t-1$.
Eq.~(\ref{eq:gp-adf marginal mean 1st attempt}) can be rewritten as $
\vec\mu_{t|t-1}^x = \E_{\vec x_{t-1}}[m_f(\vec
x_{t-1})|\vec\obs_{1:t-1}]$ with $m_f(\vec x_{t-1}) \coloneqq
\E_f[f(\vec x_{t-1})|\vec x_{t-1}]$ is the posterior mean function of
$\GP_f$.  Writing $m_f$ as a finite sum over the SE kernels centered
at all $n$ training inputs~\cite{Rasmussen2006}, the predicted mean
for each target dimension $a=1,\dotsc,D$ is
%
\begin{align}
  (\vec\mu_{t|t-1}^x)_a&=\int m_{f_a}(\vec x_{t-1})\prob(\vec
  x_{t-1}|\vec\obs_{1:t-1}) \d\vec x_{t-1}\label{eq:gp-adf time update kernel integral}\\
  &=\sum\nolimits_{i=1}^n\beta_{a_i}^x\int k_{f_a}(\vec x_{t-1},\vec
  x_i)\prob(\vec x_{t-1}|\vec\obs_{1:t-1})\d\vec x_{t-1}\,, \nonumber
\end{align}
where $\prob(\vec x_{t-1}|\vec\obs_{1:t-1}) = \gaussx{\vec
  x_{t-1}}{\vec\mu_{t-1|t-1}^x}{\mat\Sigma_{t-1|t-1}^x}$ is the filter
distribution at time $t-1$. Moreover, $\vec x_i$, $i=1,\dotsc,n$, are
the training set of $\GP_f$, $k_{f_a}$ is the covariance function of
$\GP_f$ for the $a$th target dimension (GP hyper-parameters are not
shared across dimensions), and $\vec\beta_a^x \coloneqq (\mat K_{f_a}
+ \sigma_{w_a}^2\mat I)\inv\vec y_a\in\R^n$. For dimension $a$, $\mat
K_{f_a}$ denotes the kernel matrix (Gram matrix), where $\mat
K_{f_{a_{ij}}} = k_{f_a}(\vec x_i,\vec x_j)$,
$i,j=1,\dotsc,n$. Moreover, $\vec y_a$ are the training targets, and
$\sigma_{w_a}^2$ is the learned system noise variance. The vector
$\vec\beta_a^x$ has been pulled out of the integration since it is
independent of $\vec x_{t-1}$. Note that $\vec x_{t-1}$ serves as a
test input from the perspective of the GP regression model.

For the SE covariance function in Eq.~(\ref{eq:SE kernel}), the
integral in~(\ref{eq:gp-adf time update kernel integral}) can be
computed analytically (other tractable choices are covariance
functions containing combinations of squared exponentials,
trigonometric functions, and polynomials)\@. The marginal mean is given
as
\begin{align}
(\vec\mu_{t|t-1}^x)_a&=(\vec\beta_a^x)\T\vec q_a^x
\label{eq:gp-adf time update mean}
\end{align}
where we defined
\begin{align}
 q_{a_i}^x&\coloneqq\alpha_{f_a}^2|\mat\Sigma_{t-1|t-1}^x\mat\Lambda_a\inv
  + \mat I|^{-\frac{1}{2}}\nonumber\\
&\hspace{-2mm}\times \exp\big(-\tfrac{1}{2}(\vec
  x_i-\vec\mu_{t-1|t-1}^x)\T\mat S\inv(\vec x_i-\vec\mu_{t-1|t-1}^x)\big)\,,
\label{eq:definition q}\\
\mat S &\coloneqq \mat\Sigma_{t-1|t-1}^x + \mat\Lambda_a\,,
\label{eq:S-matrix}
\end{align}
$i = 1,\dotsc,n$, being the solution to the integral in
Eq.~(\ref{eq:gp-adf time update kernel integral})\@. Here,
$\alpha_{f_a}^2$ is the signal variance of the $a$th target dimension
of $\GP_f$, a learned hyper-parameter of the SE covariance function,
see Eq.~(\ref{eq:SE kernel})\@.

\subsubsection{Marginal Covariance Matrix}
We now explicitly compute the entries of the corresponding covariance
$\mat\Sigma_{t|t-1}^x$. Using the law of total covariance, we obtain
for $a,b=1,\dotsc,D$
\begin{align}
  &(\Sigma_{t|t-1}^x)_{(ab)} = \cov_{\vec x_{t-1},f,\vec
    w}[x_t^\idx{a},x_t^\idx{b}|\vec\obs_{1:t-1}]
  \label{eq:entries marginal covariance matrix}\\
  & = \E_{\vec x_{t-1}}\big[\cov_{f,\vec w}[f_a(\vec x_{t-1})+w_a,
  f_b(\vec x_{t-1}) + w_b|\vec
  x_{t-1}]|\vec\obs_{1:t-1}\big]\nonumber\\
  &\quad + \cov_{\vec x_{t-1}}\big[\E_{f_a}[f_a(\vec x_{t-1})|\vec
  x_{t-1}],\E_{f_b}[ f_b(\vec x_{t-1})|\vec
  x_{t-1}]|\vec\obs_{1:t-1}\big]\,,
\nonumber
\end{align}
where we exploited in the last term that the system noise $\vec w$ has
mean zero. Note that Eq.~\eqref{eq:entries marginal covariance matrix}
is the sum of the covariance of (conditional) expected values and the
expectation of a (conditional) covariance. We analyze these terms in
the following.

The \emph{covariance of the expectations}  in
Eq.~\eqref{eq:entries marginal covariance matrix} is
\begin{align}
\int m_{f_a}(\vec x_{t-1}) m_{f_b}(\vec x_{t-1})\prob(\vec
x_{t-1})\d\vec x_{t-1} - (\mu_{t|t-1}^x)_a(\mu_{t|t-1}^x)_b\,,
\end{align}
where we used that $\E_f[f(\vec x_{t-1})|\vec x_{t-1}] = m_f(\vec
x_{t-1})$. With $\vec\beta_a^x = (\mat K_a + \sigma_{w_a}^2\mat
I)\inv\vec y_a$ and $m_{f_a}(\vec x_{t-1})=k_{f_a}(\mat X,\vec
x_{t-1})\T\vec\beta^x_a$, we obtain
\begin{align}
  \cov&_{\vec x_{t-1}}[m_{f_a}(\vec x_{t-1}), m_{f_b}(\vec
  x_{t-1})|\vec
  \obs_{1:t-1}]\nonumber\\
  &= (\vec \beta_a^x)\T\mL\vec\beta_b^x -
  (\mu_{t|t-1}^x)_a(\mu_{t|t-1}^x)_b
\label{eq:common term marginal covariance term}\,.
\end{align}
Following~\cite{Deisenroth2010b}, the entries of $\mat Q\in\R^{n\times
  n}$ are given as
\begin{align}
  Q_{{ij}}& = k_{f_a}(\vec x_i,\vec\mu_{t-1|t-1}^x)k_{f_b}(\vec
  x_j,\vec\mu_{t-1|t-1}^x)/\sqrt{|\mat
  R|}\label{eq: L-matrix2}\\
  &\quad\times\exp\big(\tfrac{1}{2}\vec z_{ij}\T\mat
  R\inv\mat\Sigma_{t-1|t-1}^x\vec z_{ij}\big)=
  \exp(n_{ij}^2)/\sqrt{|\mat
      R|}\,,\nonumber\\
  n_{ij}^2 & =
  \log(\alpha_{f_a}^2)+\log(\alpha_{f_b}^2)\nonumber\\
  &\quad -\tfrac{1}{2}\big(\vec\zeta_i\T\mat \Lambda_a\inv\vec\zeta_i
  + \vec\zeta_j\T\mat\Lambda_b\inv\vec\zeta_j-\vec z_{ij}\T \mat
  R\inv\mat\Sigma_{t-1|t-1}^x\vec z_{ij}\big)\,,\nonumber
\end{align}
where we defined $\mat R\coloneqq
\mat\Sigma_{t-1|t-1}^x(\mat\Lambda_a\inv+\mat\Lambda_b\inv)+\mat I$,
$\vec\zeta_i\coloneqq \vec x_i-\vec\mu_{t-1|t-1}^x $, and $\vec
z_{ij}\coloneqq\mat\Lambda_a\inv\vec\zeta_i+\mat\Lambda_b\inv\vec\zeta_j$.

The \emph{expected covariance}  in
Eq.~\eqref{eq:entries marginal covariance matrix} is given as
\begin{align}
\hspace{-2mm}\E_{\vec x_{t-1}}\big[
\cov_f[f_a(\vec x_{t-1}), f_b(\vec x_{t-1})|\vec x_{t-1}]
|\vec\obs_{1:t-1}\big]
+\delta_{ab}\sigma_{w_a}^2
\label{eq:additional term for diagonal entries}
\end{align}
since the noise covariance matrix $\mat\Sigma_w$ is
diagonal. Following our GP training assumption that different target
dimensions do not covary if the input is deterministically given,
Eq.~\eqref{eq:additional term for diagonal entries} is only non-zero
if $a=b$, i.e., Eq.~\eqref{eq:additional term for diagonal entries}
plays a role only for diagonal entries of $\mat\Sigma_{t|t-1}^x$. For
these diagonal entries ($a=b$), the expected covariance in
Eq.~\eqref{eq:additional term for diagonal entries} is
\begin{align}
\alpha_{f_a}^2 - \tr\big((\mat K_{f_a} +
\sigma_{w_a}^2\mat I)\inv\mL\big)+\sigma_{w_a}^2\,.
\label{eq:additional term for diagonal entries 2}
\end{align}
Hence, the desired marginal covariance matrix in Eq.~\eqref{eq:entries
  marginal covariance matrix} is
\begin{align}
  (\Sigma_{t|t-1}^x)_{ab} = \left\{ \begin{array}{rl}
      \text{Eq.~}\eqref{eq:common term marginal covariance
        term}+\text{Eq.~}\eqref{eq:additional term for diagonal
        entries 2} &\mbox{ if $a=b$} \\
      \text{Eq.~}\eqref{eq:common term marginal covariance term}
      &\mbox{ otherwise}
       \end{array} \right.
\label{eq:marginal covariance matrix}
\end{align}

We have now solved for the marginal distribution $\prob(\vec
x_t|\vec\obs_{1:t-1})$ in Eq.~(\ref{eq:gp-ads desired joint})\@. Since
the approximate Gaussian filter distribution $\prob(\vec x_{t-1}|
\vec\obs_{1:t-1})=\gauss{\vec
  \mu_{t-1|t-1}^x}{\mat\Sigma_{t-1|t-1}^x}$ is also known, it remains
to compute the cross-covariance $\mat\Sigma_{t-1,t|t-1}^x$ to fully
determine the Gaussian approximation in Eq.~\eqref{eq:gp-ads desired
  joint}.

\subsection{Cross-Covariance}
\label{sec:cross-covariance}
By the definition of a covariance and the system Eq.~(\ref{eq:system
  equation}), the missing cross-covariance matrix
$\mat\Sigma_{t-1,t|t-1}^x $ in Eq.~(\ref{eq:gp-ads desired joint}) is
\begin{align}
  \mat\Sigma_{t-1,t|t-1}^x &= \E_{\vec x_{t-1},f,\vec w_t}[\vec
  x_{t-1}\big(f(\vec x_{t-1}) +
  \vec w_t\big)\T|\vec\obs_{1:t-1}]\nonumber\\
  &\quad - \vec\mu_{t-1|t-1}^x(\vec \mu_{t|t-1}^x)\T\,,
\end{align}
where $\vec\mu_{t-1|t-1}^x$ is the mean of the filter update at time
step $t-1$ and $\vec\mu_{t|t-1}^x$ is the mean of the time update, see
Eq.~(\ref{eq:gp-adf marginal mean 1st attempt})\@. Note that we
explicitly average out the model uncertainty about $f$. Using the law
of total expectations, we obtain
\begin{align}
  \mat\Sigma_{t-1,t|t-1}^x
  &=\E_{\vec x_{t-1}}\big[\vec x_{t-1}\, \E_{f,\vec w_t}[f(\vec
  x_{t-1}) + \vec w_t | \vec x_{t-1}]\T
  \,|\vec\obs_{1:t-1}\big]\nonumber\\
  &\quad- \vec\mu_{t-1|t-1}^x(\vec
  \mu_{t|t-1}^x)\T\\
  &= \E_{\vec x_{t-1}}\big[\vec x_{t-1} m_f(\vec
  x_{t-1})\T|\vec\obs_{1:t-1}\big]\nonumber\\
  &\quad - \vec\mu_{t-1|t-1}^x(\vec \mu_{t|t-1}^x)\T\,,
\end{align}
where we used the fact that $\E_{f,\vec w_t}[f(\vec x_{t-1}) + \vec
w_t | \vec x_{t-1}]=m_f(\vec x_{t-1})$ is the mean function of
$\GP_f$, which models the mapping $\vec x_{t-1}\mapsto\vec x_t$, evaluated
at $\vec x_{t-1}$. We thus obtain
\begin{align}
  \mat\Sigma_{t-1,t|t-1}^x &= \int \vec x_{t-1} m_f(\vec
  x_{t-1})\T\prob(\vec x_{t-1}|\vec\obs_{1:t-1})\d\vec
  x_{t-1}\label{eq:cross-covariance integration}\\
  &\quad - \vec\mu_{t-1|t-1}^x(\vec \mu_{t|t-1}^x)\T\,.\nonumber
\end{align}
Writing $m_f(\vec x_{t-1})$ as a finite sum over
kernels~\cite{Rasmussen2006} and moving the integration into this sum,
the integration in Eq.~\eqref{eq:cross-covariance integration} turns
into
\begin{align*}
  &\quad\int \vec x_{t-1} m_{f_a}(\vec x_{t-1})\prob(\vec
  x_{t-1}|\vec\obs_{1:t-1})\d\vec x_{t-1}\nonumber \\
  &=\sum_{i = 1}^n \beta_{a_i}^x \int\vec x_{t-1} k_{f_a}(\vec
    x_{t-1},\vec x_i)\prob(\vec x_{t-1}|\vec\obs_{1:t-1})\d\vec
    x_{t-1} 
\end{align*}
for each state dimension $a = 1,\dotsc,D$.
With the SE covariance function $k_{\text{SE}}$ defined in
Eq.~(\ref{eq:SE kernel}), we compute the integral analytically and
obtain
\begin{align}
  & \int \vec x_{t-1} m_{f_a}(\vec x_{t-1})\prob(\vec
  x_{t-1}|\vec\obs_{1:t-1})\d\vec x_{t-1} 
\label{eq:input-output-cov. integral smoother}\\
  &= \sum_{i = 1}^n\beta_{a_i}^x \int \vec x_{t-1} c_3\gauss{\vec
    x_i}{\mat\Lambda_a}\gauss{\vec\mu_{t-1|t-1}^x}{\mat\Sigma_{t-1|t-1}^x}\d\vec
  x_{t-1}\nonumber\,,
\end{align}
where we defined
$c_3\inv=(\alpha_{f_a}^{2}(2\pi)^{\tfrac{D}{2}}\sqrt{|\mat\Lambda_a|})\inv$,
such that $k_{f_a}(\vec x_{t-1},\vec x_i)=c_3\gaussx{\vec
  x_{t-1}}{\vec x_i}{\mat\Lambda_a}$. In the definition of $c_3$,
$\alpha_{f_a}^2$ is a hyper-parameter of $\GP_f$ responsible for the
variance of the latent function in dimension $a$. Using the definition
of $\mat S$ in Eq.~\eqref{eq:S-matrix}, the product of the two
Gaussians in Eq.~(\ref{eq:input-output-cov. integral smoother})
results in a new (unnormalized) Gaussian $c_4\inv\gaussx{\vec
  x_{t-1}}{\vec\psi_i}{\mat\Psi}$ with
\begin{align*}
  c_4\inv&=(2\pi)^{-\tfrac{D}{2}}|\mat\Lambda_a+\mat
  \Sigma_{t-1|t-1}^x|^{-\tfrac{1}{2}}\nonumber\\
  &\quad\times\exp\big(-\tfrac{1}{2}(\vec x_i
  -\vec\mu_{t-1|t-1}^x)\T\mat S\inv(\vec x_i
  -\vec\mu_{t-1|t-1}^x)\big)\,, \nonumber\\
  \mat\Psi &= (\mat\Lambda_a\inv+(\mat \Sigma_{t-1|t-1}^x)\inv)\inv\,,\nonumber\\
  \vec\psi_i &= \mat\Psi(\mat\Lambda_a\inv\vec x_i + (\mat
  \Sigma_{t-1|t-1}^x)\inv\vec\mu_{t-1|t-1}^x)\,.\nonumber
\end{align*}
Pulling all constants outside the integral in
Eq.~(\ref{eq:input-output-cov. integral smoother}), the integral
determines the expected value of the product of the two Gaussians,
$\vec\psi_i$. For $a=1,\dotsc,D$, we obtain
\begin{align*}
  &\quad \E[\vec x_{t-1}\,
  x_{t_a}|\vec\obs_{1:t-1}]=\sum\nolimits_{i=1}^n  c_3c_4\inv\beta_{a_i}^x\vec\psi_i\,.
\end{align*}
Using $c_3c_4\inv = q_{a_i}^x$, see Eq.~(\ref{eq:definition q}), and
some matrix identities, we finally obtain
\begin{align}
  \sum\nolimits_{i=1}^n\beta_{a_i}^x
  q_{a_i}^x\mat\Sigma_{t-1|t-1}^x(\mat\Sigma_{t-1|t-1}^x +
  \mat\Lambda_a)\inv(\vec x_i-\vec\mu_{t-1|t-1}^x)
  \label{eq:joint final cross cov}
\end{align}
for $ \cov_{\vec x_{t-1},f,\vec w_t}[\vec
x_{t-1},x_{t_a}|\vec\obs_{1:t-1}] $, and the joint covariance matrix
of $\prob(\vec x_{t-1},\vec x_t|\vec\obs_{1:t-1})$ and, hence, the
full Gaussian approximation in Eq.~(\ref{eq:gp-ads desired joint}) is
completely determined.

With the mean and the covariance of the joint distribution $\prob(\vec
x_{t-1},\vec x_t|\vec\obs_{1:t-1})$ given by Eqs.~(\ref{eq:gp-adf time
  update mean}),~\eqref{eq:marginal covariance matrix},~(\ref{eq:joint
  final cross cov}), and the filter step, all necessary components are
provided to compute the smoothing distribution $\prob(\vec
x_t|\vec\obs_{1:T})$ analytically~\cite{Deisenroth2011}.

\section{Simulations}
\label{sec:results}
In the following, we present results analyzing the robustness of
state-of-the art nonlinear filters (Sec.~\ref{sec:filter robustness})
and the performances of the corresponding smoothers
(Sec.~\ref{sec:pendulum tracking})\@.

\subsection{Filter Robustness}
\label{sec:filter robustness}
We consider the nonlinear stochastic dynamic system
\begin{align}
  x_t &= \tfrac{x_{t-1}}{2} + \tfrac{25\,x_{t-1}}{1+x_{t-1}^2} +
  w_t\,,\quad w_t\sim\gauss{0}{\sigma_w^2=0.2^2}\,, \label{eq:kitagawa
    system}\\
  \obs_t &= 5\,\sin(x_t) + v_t\,,\quad
  v_t\sim\gauss{0}{\sigma_v^2=0.2^2}\,,\label{eq:kitagawa measurement}
\end{align}
which is a modified version of the model used
in~\cite{Kitagawa1996,Doucet2000}. The system is modified in two ways:
First, Eq.~(\ref{eq:kitagawa system}) does not contain a purely
time-dependent term in the system, which would not allow for learning
stationary transition dynamics. Second, we substituted a sinusoidal
measurement function for the originally quadratic measurement function
used by~\cite{Kitagawa1996} and~\cite{Doucet2000}. The sinusoidal
measurement function increases the difficulty in computing the
marginal distribution $\prob(\vec\obs_t|\vec\obs_{1:t-1})$ if the time
update distribution $\prob(\vec x_t|\vec\obs_{1:t-1})$ is fairly
uncertain: While the quadratic measurement function can only lead to
bimodal distributions (assuming a Gaussian input distribution), the
sinusoidal measurement function in Eq.~(\ref{eq:kitagawa measurement})
can lead to an arbitrary number of modes---for a broad input
distribution.

The prior variance was set to $\sigma_0^2 = 0.5^2$, i.e., the initial
uncertainty was fairly high. The system and measurement noises (see
Eqs.~(\ref{eq:kitagawa system})--(\ref{eq:kitagawa measurement})) were
relatively small considering the amplitudes of the system function and
the measurement function.
For the numerical analysis, a linear grid in the interval $[-3,3]$ of
mean values $(\mu_0^x)_i$, $i=1,\dotsc,100$, was defined. Then, a
single latent (initial) state $x_0^\idx{i}$ was sampled from
$\prob(x_0^\idx{i})=\gauss{(\mu_0^x)_i}{\sigma_0^2}$,
$i=1,\dotsc,100$.  

For the dynamic system in Eqs.~(\ref{eq:kitagawa
  system})--(\ref{eq:kitagawa measurement}), we analyzed the
robustness in a single filter step of the EKF, the UKF, the CKF, an
SIR PF (sequential importance resampling particle filter) with 200
particles, the GP-UKF, and the GP-ADF against the ground truth,
closely approximated by the Gibbs-filter~\cite{Deisenroth2011}.
Compared to the evaluation of longer trajectories, evaluating a single
filter step makes it easier to analyze the robustness of individual
filtering algorithms.

\begin{table*}[t]
\centering
  \caption{Average filter performances (RMSE, MAE, NLL) with standard
    errors ($95\%$ confidence interval) and p-values testing the
    hypothesis that the other filters are better than the GP-ADF using
    a one-sided t-test.}
\scalebox{1}{
\begin{tabular}{l || r r r}
\label{tab:evaluation nonlinear system}
  & $\rmse_x$ (p-value) & $\mae_x$ (p-value) & $\nll_x$ (p-value)\\
  \hline
  \hline
  EKF~\cite{Maybeck1979} 
  & $3.62\pm 0.212$ ($p = 4.1\times 10^{-2}$) 
  & $2.36\pm 0.176$ ($p=0.38$) 
  & $3.05\times 10^3 \pm 3.02\times 10^2$ ($p<10^{-4}$)\\ 
  UKF~\cite{Julier2004} 
  & $10.5\pm 1.08$ ($p<10^{-4}$) 
  & $8.58\pm 0.915$ ($p<10^{-4}$) 
  & $25.6 \pm 3.39$ ($p<10^{-4}$)\\ 
  CKF~\cite{Arasaratnam2009} 
  & $9.24 \pm 1.13$ ($p=2.8\times 10^{-4}$) 
  & $7.31\pm 0.941$ ($p=4.2\times 10^{-4}$) 
  & $2.22\times 10^2 \pm 17.5$ ($p<10^{-4}$)\\ 
  \hline
  GP-UKF~\cite{Ko2009} 
  & $5.36\pm 0.461$ ($p = 7.9\times 10^{-4}$) 
  & $3.84\pm 0.352$ ($p = 3.3\times 10^{-3}$) 
  & $6.02\pm 0.497$ ($p < 10^{-4}$)\\ 
  GP-ADF~\cite{Deisenroth2009a} 
  & $\mathbf{2.85 \pm 0.174}$ 
  & $\mathbf{2.17\pm 0.151}$ 
  & $\mathbf{1.97 \pm 6.55\times10^{-2}}$\\ 
  \hline
  Gibbs-filter~\cite{Deisenroth2011} 
  & {\green{$\mathbf{2.82\pm 0.171}$}} ($p=0.54$) 
  & {\green{$\mathbf{2.12\pm 0.148}$}} ($p=0.56$) 
  & {\green{$\mathbf{1.96\pm 6.62\times 10^{-2}}$}} ($p=0.55$) 
  \\
  SIR PF 
  & {\green{$\mathbf{1.57 \pm 7.66\times 10^{-2}}$}} ($p=1.0$) 
   & {\green{$\mathbf{0.36\pm 2.28\times 10^{-2}}$}}  ($p=1.0$) 
  & {\green{$\mathbf{1.03 \pm 7.30\times 10^{-2}}$}} ($p=1.0$) 
\end{tabular}
}
\end{table*}
Tab.~\ref{tab:evaluation nonlinear system} summarizes the expected
performances (RMSE: root-mean-square error, MAE: mean-absolute error,
NLL: negative log-likelihood) of the EKF, the UKF, the CKF, the
GP-UKF, the GP-ADF, the Gibbs-filter, and the SIR PF for estimating
the latent state $x$. The results in the table are based on averages
over 1{,}000 test runs and 100 randomly sampled start states per test
run (see experimental setup)\@. The table also reports the 95\% standard
error of the expected performances. The $\star$ indicates a method
developed in this paper.
Tab.~\ref{tab:evaluation nonlinear system} indicates that the GP-ADF
is the most robust filter and statistically significantly outperforms
all filters but the sampling-based Gibbs-filter and the SIR PF\@. The
green color highlights a near-optimal Gaussian filter (Gibbs-filter)
and the near-optimal particle filter. Amongst all other filters the
GP-ADF is the closest Gaussian filter to the computationally expensive
Gibbs-filter~\cite{Deisenroth2011}. Note that the SIR PF is not a
Gaussian filter and is able to express multi-modality in
distributions. Therefore, its performance is typically better than the
one of Gaussian filters. The difference between the SIR PF and a
near-optimal Gaussian filter, the Gibbs-filter, is expressed in
Tab.~\ref{tab:evaluation nonlinear system}. The performance difference
essentially depicts how much we lose by using a Gaussian filter
instead of a particle filter. The NLL values for the SIR PF are
obtained by moment-matching the particles.

\begin{figure*}[tb]
  \centering \subfigure[UKF time update $\prob(x_1|\emptyset)$, which
  misses out substantial probability mass of the true predictive
  distribution.]  {
\includegraphics[width = \twofig]{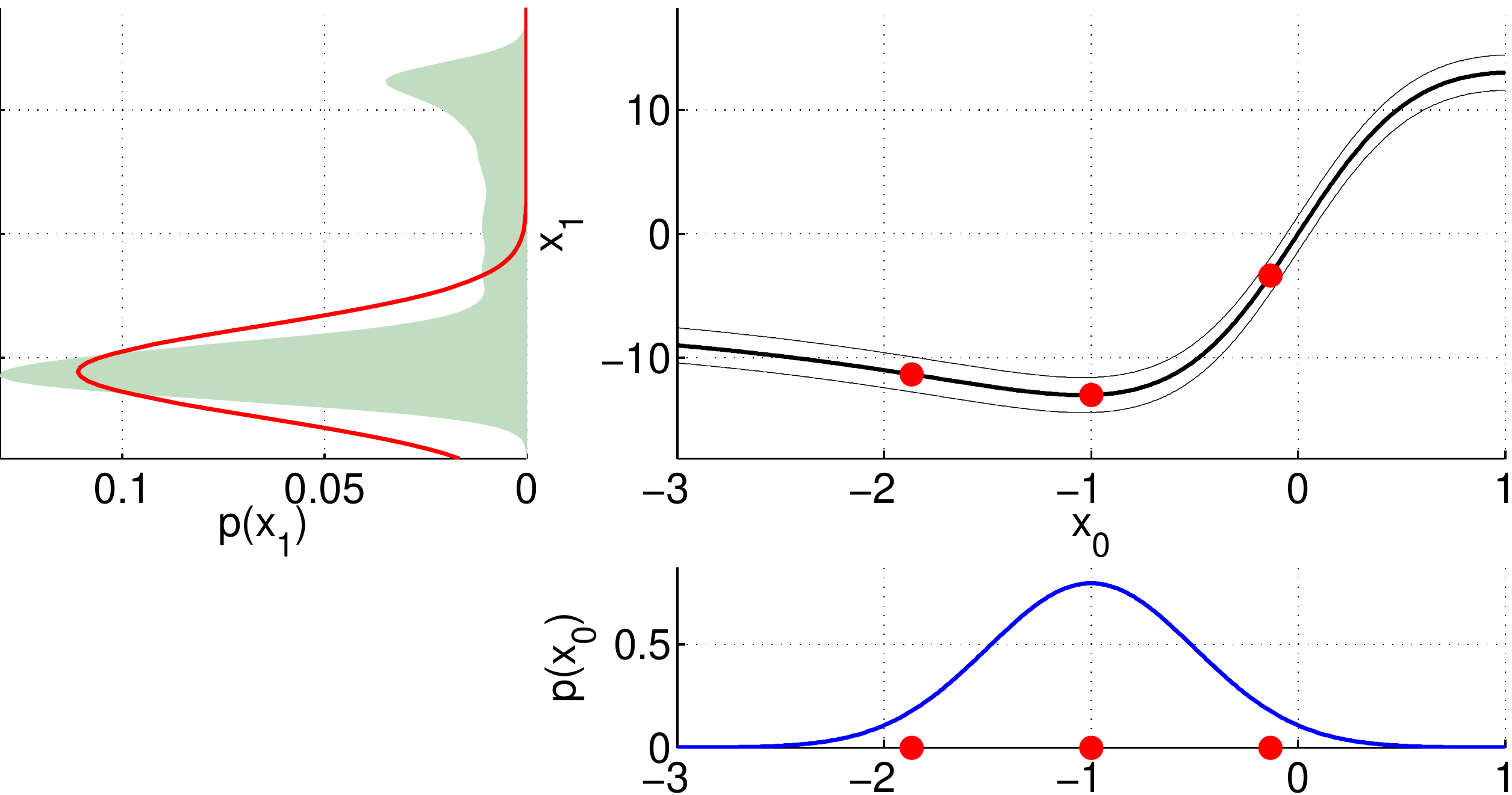}
\label{fig:ukf_fails1}
}
\hspace{1.5cm}
\subfigure[UKF determines $\prob(\obs_1|\emptyset)$, which is too
sensitive and cannot explain the actual measurement $\obs_1$ (black
dot, left sub-figure)\@.]{
\includegraphics[width=\twofig]{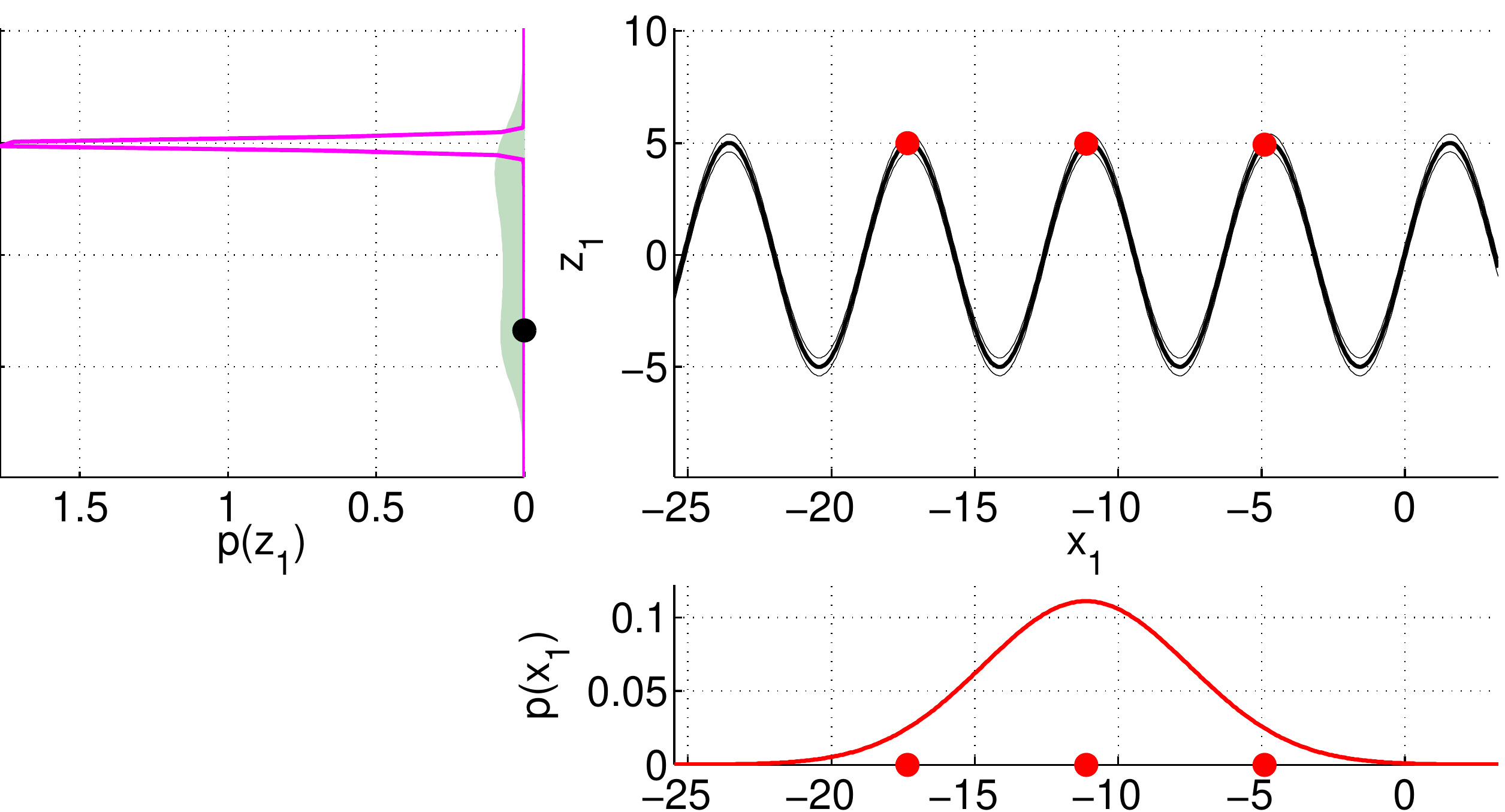}
\label{fig:ukf_fails2}
}
\caption[Degeneracy of the unscented transformation.]{Degeneracy of
  the unscented transformation (UT) underlying the UKF\@. Input
  distributions to the UT are the Gaussians in the sub-figures at the
  bottom in each panel. The functions the UT is applied to are shown
  in the top right sub-figures, i.e, the transition mapping,
  Eq.~(\ref{eq:kitagawa system}), in Panel~\subref{fig:ukf_fails1},
  and the measurement mapping, Eq.~(\ref{eq:kitagawa measurement}), in
  Panel~\subref{fig:ukf_fails2}\@. Sigma points are marked by red
  dots. The predictive distributions are shown in the left sub-figures
  of each panel. The true predictive distributions are the shaded
  areas; the UT predictive distributions are the solid Gaussians. The
  predictive distribution of the time update in
  Panel~\subref{fig:ukf_fails1} equals the input distribution at the
  bottom of Panel~\subref{fig:ukf_fails2}.}
\label{fig:ukf failing}
\end{figure*}

The poor performance of the EKF is due to linearization errors. The
filters based on small sample approximations of densities (UKF,
GP-UKF, CKF) suffer from the degeneracy of these approximations, which
is illustrated in Fig.~\ref{fig:ukf failing}. Note that the CKF uses a
smaller set of cubature points than the UKF to determine predictive
distributions, which makes the CKF statistically even less robust than
the UKF\@.
%

\subsection{Smoother Robustness}
\label{sec:pendulum tracking}

We consider a pendulum tracking example taken
from~\cite{Deisenroth2009a}. We evaluate the performances of four
filters and smoothers, the EKF/EKS, the UKF/URTSS, the
GP-UKF/GP-URTSS, the CKF/CKS, the Gibbs-filter/smoother, and the
GP-ADF/GP-RTSS\@.
The pendulum has mass $m=\unit[1]{kg}$ and length $l=\unit[1]{m}$. The
state $\vec x=[\dot\varphi,\varphi]\T$ of the pendulum is given by the
angle $\varphi$ (measured anti-clockwise from hanging down) and the
angular velocity $\dot\varphi$. The pendulum can exert a constrained
torque $u\in[-5,5]\,\unit{Nm}$.
We assumed a frictionless system such that the transition function $f$
is
\begin{align}
f(\vec x_t,u_t) & = \int_{t}^{t+\Delta_t}
\begin{bmatrix}
  \displaystyle
  \tfrac{u(\tau) -
  0.5\,mlg\sin\varphi(\tau)}{0.25\,ml^2+I}
  \\
  \dot\varphi(\tau)
\end{bmatrix}
\d\tau
\label{eq:ODE equation tracking}\,,
\end{align}
where $I$ is the moment of inertia and $g$ the acceleration of
gravity. Then, the successor state
\begin{align}
  \vec x_{t+1} & =\vec x_{t+\Delta_t} = f(\vec x_t,u_t) + \vec w_t\,,
  \label{eq:sys.Eq.pend}
\end{align}
was computed using an ODE solver for Eq.~(\ref{eq:ODE equation
  tracking}) with a zero-order hold control signal $u(\tau)$. In
Eq.~\eqref{eq:sys.Eq.pend}, we set $\mat\Sigma_w = \diag([0.5^2,
0.1^2])$.  In our experiment, the torque was sampled randomly
according to $u\sim\mathcal U[-5,5]\,\unit{Nm}$ and implemented using
a zero-order-hold controller.
Every time increment $\Delta_t=\unit[0.2]{s}$, the state was measured
according to
\begin{align}
\obs_{t} &=
\arctan\big( \tfrac{-1-l\sin(\varphi_t)}{0.5-l\cos(\varphi_t)}\big)
+v_t\,,\quad\sigma_v^2=0.05^2\label{eq:meas.Eq.pend}\,.
\end{align}
Note that the scalar measurement Eq.~\eqref{eq:meas.Eq.pend}
solely depends on the angle. Thus, the full distribution of the latent
state $\vec x$ had to be reconstructed using the cross-correlation
information between the angle and the angular velocity.

Trajectories of length $T=\unit[6]{s} = 30$ time steps were started
from a state sampled from the prior $\prob(\vec x_0) =
\gauss{\vec\mu_0}{\mat\Sigma_0}$ with $\vec\mu_0=[0, 0]\T$ and
$\mat\Sigma_0=\diag([0.01^2, (\tfrac{\pi}{16})^2])$. For each
trajectory, GP models $\GP_f$ and $\GP_g$ are learned based on
randomly generated data using either 250 or 20 data points.

Tab.~\ref{tab:pendulum results} reports the expected values of the
$\nll_x$-measure for the EKF\slash EKS, the UKF\slash URTSS, the
GP-UKF\slash GP-URTSS, the GP-ADF\slash GP-RTSS, and the CKF\slash CKS
when tracking the pendulum over a horizon of $\unit[6]{s}$, averaged
over 1{,}000 runs.
%
\begin{table*}[tb]
  \caption{Expected filtering and smoothing performances (pendulum
    tracking) with $95\%$ confidence intervals.}
\label{tab:pendulum results}
\centering
\scalebox{1}{
\begin{tabular}{c|r r r r r r}
  Filters  & EKF~\cite{Maybeck1979}& UKF~\cite{Julier2004} & CKF~\cite{Arasaratnam2009} &
  GP-UKF$_{250}$~\cite{Ko2009} & GP-ADF$_{250}$~\cite{Deisenroth2009a} &
  GP-ADF$_{20}$~\cite{Deisenroth2009a} \\
  \hline 
  $\E[\nll_x]$ 
  & $1.6\times 10^2\pm 29.1$ 
  & $6.0 \pm 3.02$ 
  & $28.5\pm 9.83$ 
  & $4.4 \pm 1.32$ 
  & $\mathbf{1.44\pm 0.117}$ 
  & $6.63 \pm 0.149$\\ 
  \hline
  \hline
  Smoothers & EKS~\cite{Maybeck1979} & URTSS~\cite{Sarkka2008} &  CKS~\cite{Deisenroth2011} &
  GP-URTSS$_{250}^\star$ & GP-RTSS$_{250}^\star$ & GP-RTSS$_{20}^\star$\\
  \hline
  $\E[\nll_x]$
  & \red{$\mathbf{3.3  \times 10^2  \pm 60.5}$} 
  & \red{$\mathbf{17.2 \pm 10.0}$} 
  & \red{$\mathbf{72.0\pm 25.1}$} 
  & \red{$\mathbf{10.3              \pm  3.85}$} 
  & $\mathbf{1.04\pm 0.204}$ 
  & $6.57 \pm 0.148$ 
\end{tabular}
}
\end{table*}
As in the example in Sec.~\ref{sec:filter robustness}, the
$\nll_x$-measure emphasizes the robustness of our proposed method: The
GP-RTSS is the only method that consistently reduced the negative
log-likelihood value compared to the corresponding filtering
algorithm. Increasing the $\nll_x$-values (red color in
Tab.~\ref{tab:pendulum results}) occurs when the filter distribution
cannot explain the latent state\slash measurement, an example of which
is given in Fig.~\ref{fig:ukf_fails2}. Even with only 20 training
points, the GP-ADF/GP-RTSS outperform the commonly used EKF/EKS,
UKF/URTSS, CKF/CKS\@.

We experimented with even smaller signal-to-noise ratios. The GP-RTSS
remains robust, while the other smoothers remain unstable. 

\section{Discussion and Conclusion}
\label{sec:discussion}
In this paper, we presented GP-RTSS, an analytic Rauch-Tung-Striebel
smoother for GP dynamic systems, where the GPs with SE covariance
functions are practical implementations of universal function
approximators. We showed that the GP-RTSS is more robust to
nonlinearities than state-of-the-art smoothers. There are two main
reasons for this: First, the GP-RTSS relies neither on linearization
(EKS) nor on density approximations (URTSS/CKS) to compute an optimal
Gaussian approximation of the predictive distribution when mapping a
Gaussian distribution through a nonlinear function. This property
avoids incoherent estimates of the filtering and smoothing
distributions as discussed in Sec~\ref{sec:filter robustness}. Second,
GPs allow for more robust ``system identification'' than standard
methods since they coherently represent uncertainties about the system
and measurement functions at locations that have not been encountered
in the data collection phase. The GP-RTSS is a robust smoother since
it accounts for model uncertainties in a principled Bayesian way.

After training the GPs, which can be performed offline, the
computational complexity of the GP-RTSS (including filtering) is
$\mathcal O(T(E^3 + n^2(D^3+E^3)))$ for a time series of length
$T$. Here, $n$ is the size of the GP training sets, and $D$ and $E$
are the dimensions of the state and the measurements, respectively.
The computational complexity is due to the inversion of the $D$ and
$E$-dimensional covariance matrices, and the computation of the matrix
$\mL\in\R^{n\times n}$ in Eq.~(\ref{eq: L-matrix2}), required for each
entry of a $D$ and $E$-dimensional covariance matrix.  The
computational complexity scales linearly with the number of time
steps.  The computational demand of classical Gaussian smoothers, such
as the URTSS and the EKS is $\mathcal O(T(D^3+E^3))$.  Although not
reported here, we verified the computational complexity
experimentally.
Approximating the online computations of the GP-RTSS by numerical
integration or grids scales poorly with increasing dimension. These
problems already appear in the histogram filter~\cite{Thrun2005}.  By
explicitly providing equations for the solution of the involved
integrals, we show that numerical integration is not necessary and the
GP-RTSS is a practical approach to filtering in GP dynamic systems.


Although the GP-RTSS is computationally more involved than the URTSS,
the EKS, and the CKS, this does not necessarily imply that smoothing
with the GP-RTSS is slower: function evaluations, which are heavily
used by the EKS\slash CKS\slash URTSS are not necessary in the GP-RTSS
(after training)\@. In the pendulum example, repeatedly calling the
ODE solver caused the EKS\slash CKS\slash URTSS to be slower than the
GP-RTSS (with 250 training points) by a factor of two.

The increasing use of GPs for model learning in robotics and control
will eventually require principled smoothing methods for GP models. To
our best knowledge, the proposed GP-RTSS is the most principled
GP-smoother since all computations can be performed analytically
exactly, i.e., without function linearization or sigma/cubature
point representation of densities, while exactly integrating out the
model uncertainty induced by the GP distribution.

Code will be made publicly available at
\texttt{\url{http://mloss.org}}.

\section*{Acknowledgements}
This work was partially supported by ONR MURI grant N00014-09-1-1052,
by Intel Labs, and by DataPath, Inc.

\ifCLASSOPTIONcaptionsoff
  \newpage
\fi


\begin{thebibliography}{10}
\providecommand{\url}[1]{#1}
\csname url@samestyle\endcsname
\providecommand{\newblock}{\relax}
\providecommand{\bibinfo}[2]{#2}
\providecommand{\BIBentrySTDinterwordspacing}{\spaceskip=0pt\relax}
\providecommand{\BIBentryALTinterwordstretchfactor}{4}
\providecommand{\BIBentryALTinterwordspacing}{\spaceskip=\fontdimen2\font plus
\BIBentryALTinterwordstretchfactor\fontdimen3\font minus
  \fontdimen4\font\relax}
\providecommand{\BIBforeignlanguage}[2]{{%
\expandafter\ifx\csname l@#1\endcsname\relax
\typeout{** WARNING: IEEEtran.bst: No hyphenation pattern has been}%
\typeout{** loaded for the language `#1'. Using the pattern for}%
\typeout{** the default language instead.}%
\else
\language=\csname l@#1\endcsname
\fi
#2}}
\providecommand{\BIBdecl}{\relax}
\BIBdecl

\bibitem{Anderson2005}
B.~D.~O. Anderson and J.~B. Moore, \emph{Optimal {Filtering}}.\hskip 1em plus
  0.5em minus 0.4em\relax Dover Publications, 2005.

\bibitem{Astrom2006}
K.~J. {\AA}str\"om, \emph{Introduction {to Stochastic Control Theory}}.\hskip
  1em plus 0.5em minus 0.4em\relax Dover Publications, Inc., 2006.

\bibitem{Thrun2005}
S.~Thrun, W.~Burgard, and D.~Fox, \emph{Probabilistic {Robotics}}.\hskip 1em
  plus 0.5em minus 0.4em\relax The MIT Press, 2005.

\bibitem{Roweis2001}
S.~T. Roweis and Z.~Ghahramani, \emph{Kalman {Filtering and Neural
  Networks}}.\hskip 1em plus 0.5em minus 0.4em\relax Wiley, 2001, ch. Learning
  {Nonlinear Dynamical Systems using the EM Algorithm}, pp. 175--220.

\bibitem{Kalman1960}
R.~E. Kalman, ``A {N}ew {A}pproach to {L}inear {F}iltering and {P}rediction
  {P}roblems,'' \emph{Transactions of the ASME---Journal of Basic Engineering},
  vol.~82, pp. 35--45, 1960.

\bibitem{Rauch1965}
H.~E. Rauch, F.~Tung, and C.~T. Striebel, ``Maximum {Likelihood Estimates of
  Linear Dynamical Systems},'' \emph{AIAA Journal}, vol.~3, pp. 1445--1450,
  1965.

\bibitem{Roweis1999}
S.~Roweis and Z.~Ghahramani, ``A {Unifying Review of Linear Gaussian Models},''
  \emph{Neural Computation}, vol.~11, no.~2, pp. 305--345, 1999.

\bibitem{Kschischang2001}
F.~R. Kschischang, B.~J. Frey, and H.-A. Loeliger, ``Factor {Graphs and the
  Sum-Product Algorithm},'' \emph{IEEE Transactions on Information Theory},
  vol.~47, pp. 498--519, 2001.

\bibitem{Pearl1988}
J.~Pearl, \emph{Probabilistic {Reasoning in Intelligent Systems: Networks of
  Plausible Inference}}.\hskip 1em plus 0.5em minus 0.4em\relax Morgan
  Kaufmann, 1988.

\bibitem{Maybeck1979}
P.~S. Maybeck, \emph{Stochastic {Models, Estimation, and Control}}.\hskip 1em
  plus 0.5em minus 0.4em\relax Academic Press, Inc., 1979, vol. 141.

\bibitem{Julier2004}
S.~J. Julier and J.~K. Uhlmann, ``Unscented {F}iltering and {N}onlinear
  {E}stimation,'' \emph{Proceedings of the IEEE}, vol.~92, no.~3, pp. 401--422,
  2004.

\bibitem{Ko2009}
J.~Ko and D.~Fox, ``G{P-BayesFilters: Bayesian Filtering using Gaussian Process
  Prediction and Observation Models},'' \emph{Autonomous Robots}, vol.~27,
  no.~1, pp. 75--90, 2009.

\bibitem{Deisenroth2009a}
M.~P. Deisenroth, M.~F. Huber, and U.~D. Hanebeck, ``Analytic {Moment-based
  Gaussian Process Filtering},'' in \emph{International Conference on Machine
  Learning}, 2009, pp. 225--232.

\bibitem{Hanebeck2003b}
U.~D. Hanebeck, ``Optimal {Filtering of Nonlinear Systems Based on Pseudo
  Gaussian Densities},'' in \emph{Symposium on System Identification}, 2003,
  pp. 331--336.

\bibitem{Arasaratnam2009}
I.~Arasaratnam and S.~Haykin, ``Cubature {Kalman Filters},'' \emph{{IEEE}
  Transactions on Automatic Control}, vol.~54, no.~6, pp. 1254--1269, 2009.

\bibitem{Sarkka2008}
S.~S\"arkk\"a, ``Unscented {Rauch-Tung-Striebel Smoother},'' \emph{IEEE
  Transactions on Automatic Control}, vol.~53, no.~3, pp. 845--849, 2008.

\bibitem{Godsill2004}
S.~J. Godsill, A.~Doucet, and M.~West, ``Monte {Carlo Smoothing for Nonlinear
  Time Series},'' \emph{Journal of the American Statistical Association},
  vol.~99, no. 465, pp. 438--449, 2004.

\bibitem{Kitagawa1996}
G.~Kitagawa, ``Monte {Carlo Filter and Smoother for Non-Gaussian Nonlinear
  State Space Models},'' \emph{Journal of Computational and Graphical
  Statistics}, vol.~5, no.~1, pp. 1--25, 1996.

\bibitem{Deisenroth2011}
M.~P. Deisenroth and H.~Ohlsson, ``A {General Perspective on Gaussian Filtering
  and Smoothing: Explaining Current and Deriving New Algorithms},'' in
  \emph{American Control Conference}, 2011.

\bibitem{MacKay1998}
D.~J.~C. MacKay, ``Introduction to {Gaussian Processes},'' in \emph{Neural
  {Networks and Machine Learning}}.\hskip 1em plus 0.5em minus 0.4em\relax
  Springer, 1998, vol. 168, pp. 133--165.

\bibitem{Rasmussen2006}
C.~E. Rasmussen and C.~K.~I. Williams, \emph{Gaussian {Processes for Machine
  Learning}}.\hskip 1em plus 0.5em minus 0.4em\relax The MIT Press, 2006.

\bibitem{Nguyen-Tuong2009}
D.~Nguyen-Tuong, M.~Seeger, and J.~Peters, ``Local {Gaussian Process Regression
  for Real Time Online Model Learning},'' in \emph{Advances in Neural
  Information Processing Systems}, 2009, pp. 1193--1200.

\bibitem{Murray-Smith2003}
R.~Murray-Smith, D.~Sbarbaro, C.~E. Rasmussen, and A.~Girard, ``Adaptive,
  {Cautious, Predictive Control with Gaussian Process Priors},'' in
  \emph{Symposium on System Identification}, 2003.

\bibitem{Kocijan2003}
J.~Kocijan, R.~Murray-Smith, C.~E. Rasmussen, and B.~Likar, ``Predictive
  {C}ontrol with {G}aussian {P}rocess {M}odels,'' in \emph{IEEE Region 8
  Eurocon 2003: Computer as a Tool}, 2003, pp. 352--356.

\bibitem{Deisenroth2011b}
M.~P. Deisenroth, C.~E. Rasmussen, and D.~Fox, ``Learning {to Control a
  Low-Cost Manipulator using Data-Efficient Reinforcement Learning},'' in
  \emph{Robotics: Science \& Systems}, 2011.

\bibitem{Deisenroth2011c}
M.~P. Deisenroth and C.~E. Rasmussen, ``P{ILCO: A Model-Based and
  Data-Efficient Approach to Policy Search},'' in \emph{International
  Conference on Machine Learning}, 2011.

\bibitem{Atkeson1997a}
C.~G. Atkeson and J.~C. Santamar\'{\i}a, ``A {Comparison of Direct and
  Model-Based Reinforcement Learning},'' in \emph{International Conference on
  Robotics and Automation}, 1997.

\bibitem{Kern2000}
J.~Kern, ``Bayesian {Process-Convolution Approaches to Specifying Spatial
  Dependence Structure},'' Ph.D. dissertation, Institue of Statistics and
  Decision Sciences, Duke University, 2000.

\bibitem{Quinonero-Candela2003a}
J.~{Qui{\~n}onero-Candela}, A.~Girard, J.~Larsen, and C.~E. Rasmussen,
  ``Propagation {of Uncertainty in Bayesian Kernel Models---Application to
  Multiple-Step Ahead Forecasting},'' in \emph{International Conference on
  Acoustics, Speech and Signal Processing}, 2003, pp. 701--704.

\bibitem{Deisenroth2010b}
M.~P. Deisenroth, \emph{Efficient {Reinforcement Learning using Gaussian
  Processes}}.\hskip 1em plus 0.5em minus 0.4em\relax KIT Scientific
  Publishing, 2010, vol.~9.

\bibitem{Doucet2000}
A.~Doucet, S.~J. Godsill, and C.~Andrieu, ``On {Sequential Monte Carlo Sampling
  Methods for Bayesian Filtering},'' \emph{Statistics and Computing}, vol.~10,
  pp. 197--208, 2000.

\end{thebibliography}
\end{document}